\documentclass[aps,prl,reprint]{revtex4-2}

\usepackage{graphicx}
\usepackage{amsmath}
\usepackage{amssymb}
\usepackage{amsfonts}
\usepackage{dsfont}
\usepackage{hyperref}
\usepackage{color}

\begin{document}

\title{Phase Transitions in Quantum Many-Body Scars}

\author{Peter Græns Larsen}
\affiliation{Department of Physics and Astronomy, Aarhus University, DK-8000 Aarhus C, Denmark}

\author{Anne E. B. Nielsen}
\affiliation{Department of Physics and Astronomy, Aarhus University, DK-8000 Aarhus C, Denmark}

\begin{abstract}
We propose a type of phase transition in quantum many-body systems, which occurs in highly excited quantum many-body scar states, while most of the spectrum is largely unaffected. Such scar state phase transitions can be realized by embedding a matrix product state, known to undergo a phase transition, as a scar state into the thermal spectrum of a parent Hamiltonian. We find numerically that the mechanism for the scar state phase transition involves the formation or presence of low-entropy states at energies similar to the scar state in the vicinity of the phase transition point.
\end{abstract}

\maketitle

The notion of quantum phases provides a framework for describing quantum systems based on key properties, and understanding phase transitions is hence of central importance to quantum physics \cite{Sachdev_1999}. In a standard quantum phase transition the system's ground state undergoes an irregular evolution as a parameter $g$ of the model passes a phase transition point. This can happen, e.g.\ because the two lowest energies in the spectrum cross, such that the first excited state turns into the ground state and the derivative of the ground state energy with respect to $g$ is discontinuous at the phase transition point in the thermodynamic limit. Another possibility is a matrix product state (MPS) phase transition, in which the derivative of the ground state energy is continuous, while the energy gap still closes at the phase transition point \cite{PhysRevLett.97.110403}. In a ground state phase transition, it is usually only the first few excited states that are impacted significantly by the phase transition. There are, however, exceptions, e.g.\ a tower of highly excited scar states that transform into low-lying excited states as a critical point of a ground state phase transition is approached \cite{PhysRevB.105.125123,PhysRevB.106.214311}. Other systems have been shown to undergo a phase transition between a scarred phase and a non-scarred phase \cite{PhysRevB.109.L220301}.

Another type of quantum phase transition is the excited state phase transition \cite{CAPRIO20081106,Cejnar_2021}. In an excited state phase transition there is an irregular behavior of states near some phase transition energy $E(g)$ which depends on a control parameter $g$. $E(g)$ defines a line in the $(g,E)$ plane which separates different phases and a phase transition can occur from changing $g$ or changing the energy.

Here, we propose and demonstrate the existence of a different type of phase transition that we call a scar state phase transition. In this type of phase transition a single, highly excited eigenstate undergoes a phase transition when $g$ is varied, while most of the spectrum is largely unaffected by the phase transition. This is motivated, in part, by the recent interest in quantum many-body scars (QMBS). QMBSs are highly excited non-thermal energy eigenstates in otherwise thermal systems, and they are typically characterized by low entanglement entropy compared to states of similar energy \cite{Turner2018,Serbyn2021,Moudgalya_2022}. Scar states have been observed experimentally in a 1D chain of Rydberg atoms where they resulted in periodic revivals of certain initial states \cite{Bernien2017}. QMBSs appear in many different models such as the Hirsch-Hubbard model \cite{PhysRevB.102.075132}, the spin 1 XY model \cite{PhysRevLett.123.147201}, and the Affleck-Kennedy-Lieb-Tasaki model \cite{PhysRevB.101.195131} to name a few. 

As QMBS states share important properties with ground states of local Hamiltonians, one could imagine that they can undergo phase transitions. On the other hand, it is not clear how they would do this. In ground state phase transitions, associated with an energy gap closing, the ground state properties change through exchange with the low-lying excited states. Scar states are surrounded by thermal states in the spectrum, and although they are not isolated from the thermal states by an energy gap, they are isolated by an entropy gap, and it is not clear how the thermal states can alter the properties of the scar state. The underlying mechanism is hence of a different nature than for ground state phase transitions, which shows that the problem is highly nontrivial.

We achieve a scar state phase transition by choosing a low-entropy state, which is known to undergo a phase transition as a function of a parameter, and embedding it as a scar state into a parent Hamiltonian. This is achieved using the framework presented in \cite{PhysRevB.102.085120} which constructs a local Hamiltonian with a given MPS as a zero energy eigenstate. This MPS can be embedded as either a scar or ground state depending on the energy of the rest of the spectrum. It is also possible to introduce parameters, and this framework thus allows for the construction of models featuring a scar state phase transition or a ground state phase transition providing a straightforward way to compare the two.

Through numerical investigations, we identify two different ways in which the scar state phase transition can happen. One possibility is that a few low-entropy states form in the spectrum in the vicinity of the scar state near the phase transition point. Another possibility is that the spectrum already features a few additional low-entropy states, and that these states approach the energy of the scar state close to the phase transition point. This should be compared to a ground state phase transition, in which the states close to the ground state in the spectrum naturally have low entropy. The difference between the ground state and scar state phase transitions thus lies in the nature of the neighboring states in the spectrum.

\textit{MPS embedding}---Consider a MPS \cite{perezgarcia2007matrix}
\begin{equation}\label{eq:wave function} |\psi_{\text{MPS}}\rangle=C\sum_{s_1,\dots,s_L}\text{tr}\left(A^{s_1}A^{s_2}\dots A^{s_L}\right)|s_1s_2\dots s_L\rangle
\end{equation}
of a chain with $L$ sites and periodic boundary conditions. Here, $C$ is a normalization constant, $s_i$ labels the $d$ states on site $i$, and the $A^{s_i}$ are $\chi\times\chi$ matrices that define the MPS. If the bond dimension $\chi$ is sufficiently low, one can construct a set of local, Hermitian operators $h_i$ that annihilate the MPS, i.e.\ $h_i|\psi_{\text{MPS}}\rangle=0$ \cite{Fernandez-Gonzalez2015}. These operators can be combined into a Hamiltonian for which the MPS is a zero energy eigenstate
\begin{gather}
	H=\sum_{i=1}^Lh_i,\quad H|\psi_{\text{MPS}}\rangle=0.
\end{gather}
If the system is translationally invariant, the $L$ annihilation operators will be related by lattice translations. Each of the annihilation operators $h_i$ acts on $D$ sites, where $D$ must be chosen such that
\begin{gather}\label{eq:D condition}
	d^D>\chi^2.
\end{gather}
The local Hilbert space on $D$ sites has dimension $d^D$ and can be split into a subspace $\mathcal{A}$ and its complement $\mathcal{A}^C$. Locally the subspace $\mathcal{A}$ fully contains the MPS and is constructed as \cite{PhysRevB.102.085120}
\begin{gather}\label{eq:subspace}		
	\mathcal{A}=\text{span}\left\{\sum_{s_1,\dots,s_D}\text{tr}\left(XA^{s_1}\dots A^{s_D}\right)|s_1,\dots,s_D\rangle\right\},
\end{gather}
where $X$ is a complete basis of $\chi\times\chi$ matrices. The dimension of $\mathcal{A}$ is $\chi^2$ corresponding to one state for each matrix in $X$. Equation \eqref{eq:D condition} thus ensures that $\mathcal{A}^C$ is non-empty. Since the MPS is locally fully contained in $\mathcal{A}$, any $D$-local operator constructed only from states in $\mathcal{A}^C$ will annihilate the MPS by construction. The annihilation operators thus take the general form
\begin{gather}\label{eq:hi}
	h_i=\sum_{n,m\in\mathcal{A}^C}c_{nm}|\psi_n\rangle\langle\psi_m|,
\end{gather}
where $|\psi_{n}\rangle$ and $|\psi_{m}\rangle$ are states in the local subspace $\mathcal{A}^C$ and the coefficients $c_{nm}$ can be chosen freely under the condition $c_{nm}=c_{mn}^*$ which ensures that $h_i$ is Hermitian.

If the coefficients $c_{nm}$ are chosen such that $h_i$ is positive semi-definite, then the spectrum will only have non-negative energies, and the MPS, embedded at zero energy, will be a ground state. As pointed out in \cite{PhysRevB.102.085120}, however, if this is not the case, the MPS may be a highly excited state as the spectrum may contain negative energies. If the embedded MPS has low bond dimension it will also have low entanglement entropy and thus qualify as a scar state. The same MPS can thus be studied as either a scar state or a ground state depending on the choice of $c_{nm}$. For simplicity we choose the coefficients $c_{nm}$ of the two different types of embedding as
\begin{itemize}
\setlength\itemsep{0pt}
\item[\boldmath$\cdot$] Scar state embedding: $c_{nm}=\left(-1\right)^n\delta_{nm}$
\item[\boldmath$\cdot$] Ground state embedding: $c_{nm}=\delta_{nm}$
\end{itemize}
where $\delta_{nm}$ is the Kronecker delta.

One can easily introduce parameters in the above construction by letting the matrices $A^{s_i}$ depend on parameters. This feature makes this method of embedding scar states particularly suitable for constructing models with a potential scar state phase transition. Other types of embeddings, such as the Shiraishi-Mori embedding \cite{PhysRevLett.119.030601}, should work as well provided one can similarly generalize the embedding to include a control parameter.

\begin{figure}
	\includegraphics[width=\linewidth]{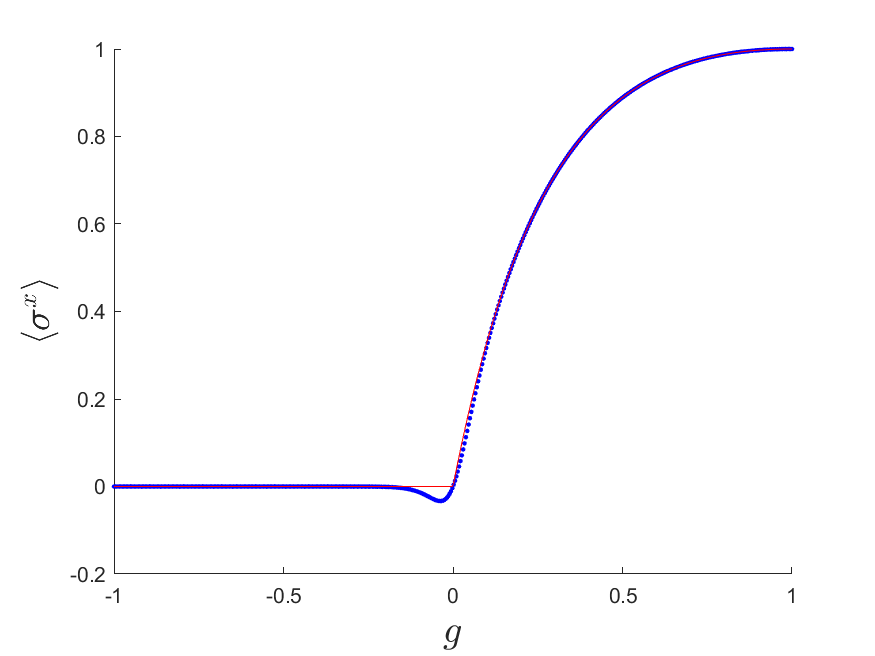}
	\caption{$\langle\sigma^x\rangle=\frac{1}{L}\langle\psi_{\text{MPS}}|\sum_{i=1}^L\sigma_i^x|\psi_{\text{MPS}}\rangle$ as a function of $g$ for infinite system size (red line) \cite{PhysRevLett.97.110403} and system size $L=18$ (blue points). The derivative of $\langle\sigma^x\rangle$ is only discontinues at $g=0$ for infinite system size, but for finite system size $\langle\sigma^x\rangle$ still changes from zero to a non-zero value over a small $g$ interval near $g=0$.}
	\label{fig:sigmax}
\end{figure}

\begin{figure*}
\includegraphics[width=\linewidth]{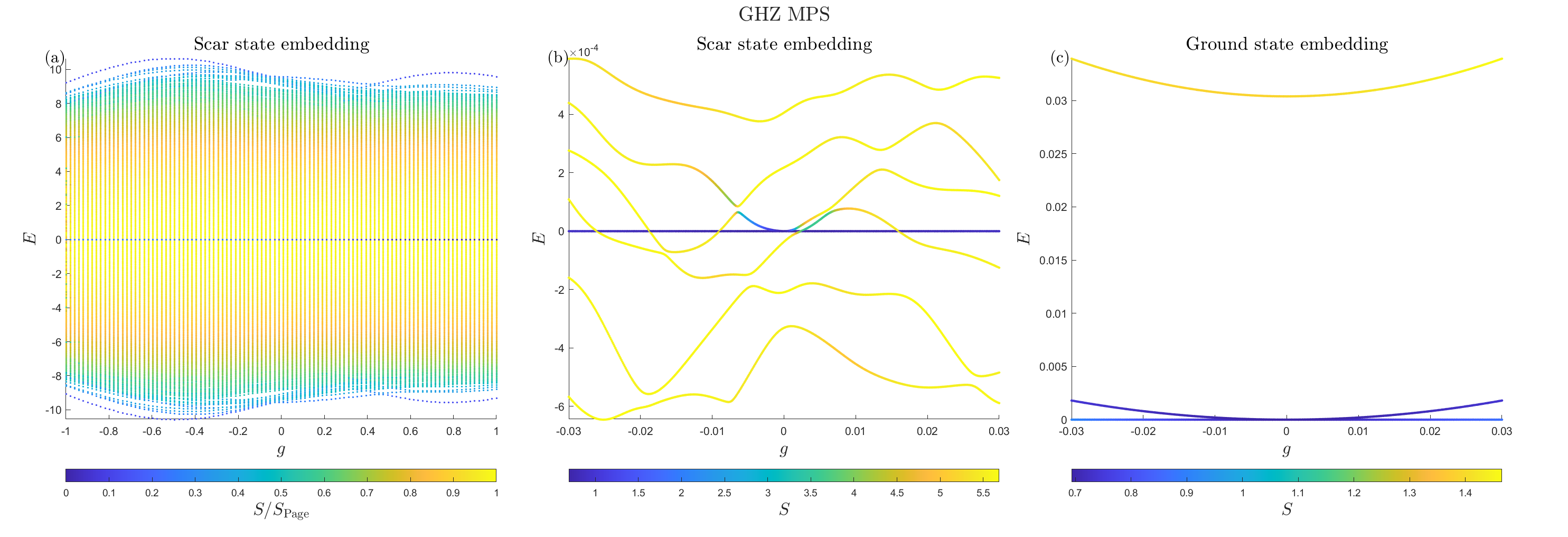}
\caption{(a) Spectrum of the zero momentum sector of the scar model obtained from the GHZ MPS with $L=18$ sites. The color shows the half-chain von Neumann entanglement entropy $S$ of the eigenstates compared to the expected entropy $S_{\textrm{Page}}$ \cite{PhysRevLett.71.1291} for a thermal state in the middle of the spectrum. The GHZ MPS is seen as a low-entropy state at zero energy in the middle of the thermal spectrum and is therefore a scar state. (b) Zoom of the same spectrum near the phase transition point at $g=0$. While most of the states in the thermal background are largely unaffected by the phase transition, a few of the states develop into low-entropy states near the phase transition point and thus get close to the scar state with respect to both energy and entropy. (c) Zoom of the spectrum of the ground state model obtained from the GHZ MPS for comparison.}
\label{fig:GHZ}
\end{figure*}

\textit{Models}---We study two translationally invariant, spin-1/2 models that both undergo a phase transition as a parameter $g$ passes through zero. Both states have bond dimension $\chi=2$ and have $d=2$ degrees of freedom on each site, so Eq.\ \eqref{eq:D condition} gives a lower bound of $D=3$. In both cases we choose the basis of matrices $X=\left\{\frac{1}{\sqrt{2}}\mathds{1},\sigma^+,\sigma^-,\frac{1}{\sqrt{2}}\sigma^z\right\}$, where $\sigma^{\pm}=\frac{1}{2}(\sigma^x\pm i\sigma^y)$ and $\sigma^b$, $b\in\{x,y,z\}$, are the Pauli matrices. For all numerical investigations below, we consider chains with $L=18$ sites. The considered MPSs belong to the zero momentum sector of their respective models so only these momentum sectors are studied.

The first model is constructed from the MPS \eqref{eq:wave function} with matrices
	\begin{gather}\label{eq:matricesGHZ}
		A^{\downarrow}=\begin{pmatrix}
			0 & 0\\
			1 & 1
		\end{pmatrix},\quad
		A^{\uparrow}=\begin{pmatrix}
			1 & g\\
			0 & 0
		\end{pmatrix}.
	\end{gather}
This MPS was shown in \cite{PhysRevLett.97.110403} to exhibit a phase transition at $g=0$. The correlation function of $m$ observables $S_i$, acting on sites $i\in\{1,2,\ldots,m\}$ or a translation thereof, is
\begin{gather}
		\langle S_1\dots S_m\rangle=\frac{\text{tr}\left[E_{\mathds{1}}^{L-m}E_{S_1}\cdots E_{S_m}\right]}{\text{tr}\left[E_{\mathds{1}}^{L}\right]},\\
		E_{S}=\sum_{s\in\{{\downarrow},{\uparrow}\}} \sum_{s'\in\{{\downarrow},{\uparrow}\}} \langle s'|S|s\rangle A^{s} \otimes (A^{s'})^*.
\end{gather}
In the thermodynamic limit ($L\rightarrow \infty$) the correlation function becomes
	\begin{gather}
		\langle S_1\dots S_m\rangle=\frac{\langle l|E_{S_1}\cdots E_{S_m}|r\rangle}{\nu_1^m},
	\end{gather}
where $|l\rangle$ and $|r\rangle$ are the left and right eigenstates of the transfer matrix $E_{\mathds{1}}$ with the largest magnitude eigenvalue $\nu_1$ \cite{PhysRevLett.97.110403}. The MPS phase transition point can thus be identified by the crossing of the two largest magnitude eigenvalues of $E_{\mathds{1}}$. This crossing often results in a discontinuity in local observables or their derivatives, while the matrices $A^{\uparrow,\downarrow}$, and therefore also the energy of the state, remain analytic.

The derivative of $\langle\sigma^x\rangle=\langle\psi_{\text{MPS}}|\sum_{i=1}^L\sigma_i^x|\psi_{\text{MPS}}\rangle/L$ shows such a discontinuity in the thermodynamic limit, where \cite{PhysRevLett.97.110403}
	\begin{gather}\nonumber
		\langle\sigma^x\rangle=0\text{ for }g<0,\\ \label{eq:sigmax}
		\langle\sigma^x\rangle=\frac{4g}{1+g^2}\text{ for }g\geq 0.
	\end{gather}
In Fig.\ \ref{fig:sigmax}, we compare $\langle\sigma^x\rangle$ for infinite system size and for system size $L=18$. The discontinuity is only exact in the thermodynamic limit, but even at finite system size $\langle\sigma^x\rangle$ changes abruptly near $g=0$. So even though the phase transition is only strictly occurring in the thermodynamic limit, there is still a visible effect at moderate system sizes.

At particular values of $g$ this MPS equals certain well known states \cite{PhysRevLett.97.110403}. At $g=-1$, the MPS equals the cluster state, at $g=0$ it equals the Greenberger–Horne–Zeilinger (GHZ) state, and at $g=1$ it is fully polarized in the $x$-direction. We will refer to it as the GHZ MPS since it equals the GHZ state at the phase transition point. For this MPS we have found an analytic form of the four local states comprising $\mathcal{A}^C$ which, up to normalization, are
	\begin{subequations}
		\begin{gather}
			|\psi_1\rangle=-g|{\uparrow}{\uparrow}{\uparrow}\rangle-|{\uparrow}{\uparrow}{\downarrow}\rangle+|{\uparrow}{\downarrow}{\uparrow}\rangle+|{\uparrow}{\downarrow}{\downarrow}\rangle,
		\end{gather}
		\begin{gather}
			|\psi_2\rangle=-|{\downarrow}{\uparrow}{\uparrow}\rangle-|{\downarrow}{\uparrow}{\downarrow}\rangle+|{\downarrow}{\downarrow}{\uparrow}\rangle+g|{\downarrow}{\downarrow}{\downarrow}\rangle,
		\end{gather}
		\begin{gather}
			\nonumber
			|\psi_3\rangle=-ga|{\uparrow}{\uparrow}{\uparrow}\rangle
			+\frac{a}{2}\left(1+g^2\right)|{\uparrow}{\uparrow}{\downarrow}\rangle
			+a|{\uparrow}{\downarrow}{\uparrow}\rangle\\ \nonumber
			-\frac{a}{2}\left(1+g^2\right)|{\uparrow}{\downarrow}{\downarrow}\rangle
			-\frac{1-a}{2}\left(1+g^2\right)|{\downarrow}{\uparrow}{\uparrow}\rangle
			+\left(1-a\right)|{\downarrow}{\uparrow}{\downarrow}\rangle\\
			+\frac{1-a}{2}\left(1+g^2\right)|{\downarrow}{\downarrow}{\uparrow}\rangle
			-g\left(1-a\right)|{\downarrow}{\downarrow}{\downarrow}\rangle,
		\end{gather}
		\begin{gather}
			\nonumber
			|\psi_4\rangle=-g\left(1-a\right)|{\uparrow}{\uparrow}{\uparrow}\rangle
			+\frac{1-a}{2}\left(1+g^2\right)|{\uparrow}{\uparrow}{\downarrow}\rangle\\ \nonumber
			+\left(1-a\right)|{\uparrow}{\downarrow}{\uparrow}\rangle
			-\frac{1-a}{2}\left(1+g^2\right)|{\uparrow}{\downarrow}{\downarrow}\rangle
			+\frac{a}{2}\left(1+g^2\right)|{\downarrow}{\uparrow}{\uparrow}\rangle\\
			-a|{\downarrow}{\uparrow}{\downarrow}\rangle
			-\frac{a}{2}\left(1+g^2\right)|{\downarrow}{\downarrow}{\uparrow}\rangle
			+ga|{\downarrow}{\downarrow}{\downarrow}\rangle,
		\end{gather}
	\end{subequations}
where $a$ is a free parameter between zero and one. The subspace simplifies for $a$ equal to zero or one, but at these values there is a degeneracy at zero energy which would have to be resolved to properly study the MPS. Instead we choose the small value $a=0.009$ which is enough to lift the degeneracy. 	

We also consider the MPS \eqref{eq:wave function} with matrices
	\begin{gather}\label{eq:matricesZ2}
		A^{\downarrow}=\begin{pmatrix}
			0 & 1\\
			0 & 0
		\end{pmatrix},\quad
		A^{\uparrow}=\begin{pmatrix}
			\sqrt{g} & 0\\
			1 & 0
		\end{pmatrix}.
	\end{gather}
This MPS has a phase transition at $g=0$. This can be shown from the eigenvalues of the transfer matrix and can also be detected from quantum fidelity \cite{PhysRevB.76.104420}. At the phase transition point $g=0$ this MPS is the $\mathbb{Z}_2$ symmetric state $|{\uparrow}{\downarrow}{\uparrow}{\downarrow}\dots\rangle+ |{\downarrow}{\uparrow}{\downarrow}{\uparrow}\dots\rangle$, and we will therefore refer to it as the $\mathbb{Z}_2$ MPS. For this MPS we have not found an analytic form of the complement subspace $\mathcal{A}^C$ and instead we have determined it numerically at each value of $g$. The Hamiltonian depends on the choice of basis for $\mathcal{A}^C$, and to get a Hamiltonian that varies continuously with $g$, we should choose a basis that varies continuously with $g$. We therefore start at $g=-1$ and iteratively do a basis transformation for each increment $\delta g$ in $g$, such that the basis choice at $g+\delta g$ resembles the basis choice at $g$ as much as possible. Specifically, we represent the basis vectors of $\mathcal{A}^C$ as an $8\times 4$ matrix $\mathcal{A}_g^C$ where each column represents a basis state. In this notation our choice of basis change is the unitary operator $U$ that minimizes $||\mathcal{A}_g^C-\mathcal{A}_{g+\delta g}^{C}U||_F$ with $||\cdot||_F$ being the Frobenius norm. This minimization problem is known as the ``Orthogonal Procrustes problem'' and has a known solution.

\begin{figure*}
\includegraphics[width=\linewidth]{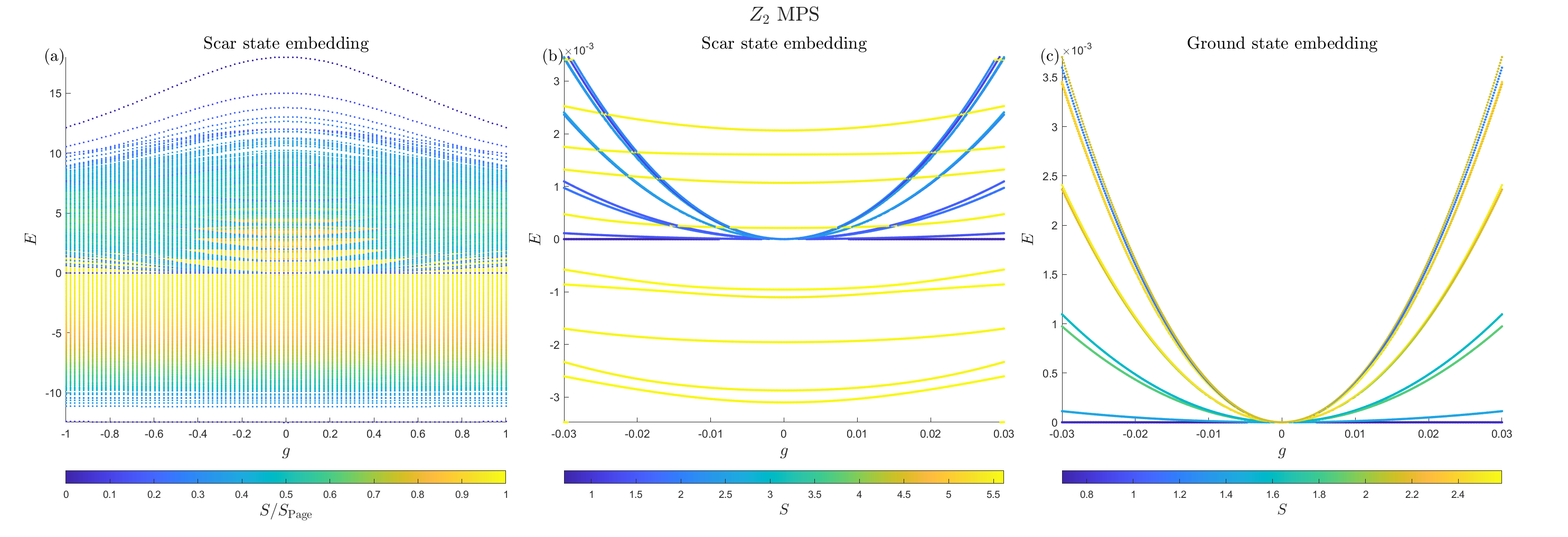}
\caption{(a) Spectrum of the zero momentum sector of the scar model obtained from the $Z_2$ MPS with $L=18$ sites. The color shows the half-chain von Neumann entanglement entropy $S$ of the eigenstates compared to the expected value $S_{\textrm{Page}}$ \cite{PhysRevLett.71.1291} for a thermal state in the middle of the spectrum. The $\mathbb{Z}_2$ MPS is a low-entropy state fixed at zero energy. There are also other low-entropy states in the spectrum, but the majority of the states are thermal and the $\mathbb{Z}_2$ MPS thus qualify as a scar state making the phase transition of this model a scar state phase transition. (b) Zoom of the same spectrum near the phase transition point at $g=0$. There are ten non-thermal states which approach the energy of the $\mathbb{Z}_2$ MPS at $g=0$. (c) Zoom of the spectrum of the ground state model obtained from the $\mathbb{Z}_2$ MPS for comparison. The first ten excited states are seen to approach the ground state at $g=0$.}
\label{fig:Z2}
\end{figure*}

\textit{Scar state phase transition}---To construct and investigate scar state phase transitions, we must choose the Hamiltonians such that the MPSs are in the middle of a thermal spectrum. That this is indeed the case for the choice made above for the model obtained from the GHZ MPS can be seen from Fig.\ \ref{fig:GHZ}(a), where the MPS appears as a low-entropy scar state surrounded by high-entropy thermal states in the middle of the spectrum.

We are primarily interested in the region close to the phase transition point, which is shown in Fig.\ \ref{fig:GHZ}(b). As $g$ varies the thermal states near zero energy exhibit avoided crossings with each other. Some of the thermal states cross the GHZ MPS without hybridizing with it, which happens because the MPS is always an exact eigenstate. Close to the phase transition, one of the eigenstates closes its energy gap to the embedded MPS, without crossing it, and reaches a similar entropy. A non-thermal energy eigenstate thus emerges from the otherwise thermal spectrum close to the scar state phase transition. In the ground state embedding of the GHZ MPS, Fig.\ \ref{fig:GHZ}(c), the first excited state closes its energy gap to the GHZ MPS and also has a similar entropy at $g=0$. The two embeddings thus share the feature of a single energy eigenstate closing its energy and entropy gap to the GHZ MPS at $g=0$. In the scar state embedding, however, this other energy eigenstate is only non-thermal close to the scar state phase transition.

The $\mathbb{Z}_2$ MPS is similarly a low-entropy state fixed at zero energy, but in this model several other states near zero energy also have low entanglement entropy as seen in Fig.\ \ref{fig:Z2}(a). The vast majority of the states near zero energy are however still thermal so the $\mathbb{Z}_2$ MPS in the scar state embedding still qualifies as a scar state. Its phase transition is therefore also a scar state phase transition in this model.

In the scar state embedding of the $\mathbb{Z}_2$ MPS, Fig.\ \ref{fig:Z2}(b), there is again a background of thermal states. There are however also ten energy eigenstates with lower entanglement entropy which have energy crossings with the thermal states like the GHZ MPS. All of these non-thermal states close their energy gap to the $\mathbb{Z}_2$ MPS at $g=0$. In the ground state embedding of the $\mathbb{Z}_2$ MPS, Fig.\ \ref{fig:Z2}(c), there are also ten energy eigenstates with low entropy which close their energy gap to the embedded MPS. The structure of the states near zero energy is thus similar between the two embeddings of the $\mathbb{Z}_2$ MPS except that the scar state embedding features a background of thermal states.

While the details of the two scar state phase transitions differ, the effect of the phase transition is limited to a few low-entropy states near the energy of the respective embedded MPS, while most of the spectrum is largely unaffected. This distinguishes the scar state phase transition from the usual excited state phase transitions in which the different phases apply to many excited states \cite{CAPRIO20081106,Cejnar_2021}. It also differs from a MPS phase transition in a ground state where the first few excited states have low entropy. Particularly, in the GHZ model the emergence of a low-entropy state from the otherwise thermal spectrum near the phase transition point contrasts the usual ground state phase transition.

\textit{Perturbations}---We note that perturbations of the $h_i$ that lie within the space spanned by the vectors in $\mathcal{A}^C$ will not affect the physics significantly, as the same MPS will remain an exact eigenstate. For other perturbations, we expect the physics to be robust if the perturbations are small enough that the MPS still has a high overlap with one of the eigenstates, similarly to the observations made in \cite{PRXQuantum.5.020365}. In particular, larger perturbations can be tolerated for smaller system sizes, which is useful since moderate size synthetic matter is a more relevant platform for implementation than condensed matter systems. An interesting next step would be to look for models, in which the space $\mathcal{A}^C$ contains all the most naturally occurring perturbations for a given experimental setup.

\textit{Conclusion}---We have proposed and demonstrated the existence of quantum scar state phase transitions. In such transitions, a highly-excited quantum many-body scar state undergoes a phase transition as a function of an external parameter, while most of the spectrum is largely unaffected. We have shown that models displaying scar state phase transitions can be constructed by embedding a MPS with low bond dimension into the middle of a thermal spectrum, where the MPS shows a phase transition as a function of a parameter $g$.

By analyzing two such models, we have identified two ways in which the transition can happen. In one model, a few thermal states transform into low-entropy states in the vicinity of the phase transition point. This is quite different from a standard ground state phase transition, in which the states near the ground state in the spectrum have low entropy even far from the transition point. In the other model, there are several low-entropy states in the spectrum, and they approach the energy of the MPS as the phase transition point is approached.

The results presented here are of fundamental interest as a different type of phase transition. The methods applied in this work are rather general and allow a large variety of models with scar state phase transitions to be constructed. This provides a starting point for further investigations. In particular, it would be interesting to see if other mechanisms for scar state phase transitions exist than the two found here. Additionally, in the framework of \cite{PhysRevB.102.085120} a tower of quasi particle scar states can potentially be constructed on top of the original MPS, which may be a way of extending the notion of the scar state phase transition to a tower of scar states.

\begin{acknowledgments}
This work has been supported by Carlsbergfondet under Grant No.\ CF20-0658.
\end{acknowledgments}

\bibliography{./bibfile}

\end{document}